\documentclass[prd,preprintnumbers,nofootinbib,aps,9pt]{revtex4}
\usepackage{subeqnarray}
\usepackage{epsfig,amsmath,amssymb}
\usepackage{mathrsfs}
\usepackage[usenames,dvipsnames]{color}
\usepackage[pagebackref=true, colorlinks=true]{hyperref}
\definecolor{redish}{rgb}{0.7,0.2,0.0}  
\definecolor{bluish}{rgb}{0.2,0.5,0.8}
\hypersetup{linkcolor=redish,          
                  citecolor=blue,        
                  filecolor=magenta,      
                  urlcolor=bluish}          
\DeclareFontFamily{U}{rsfs}{}         
\DeclareFontShape{U}{rsfs}{m}{n}{<5> rsfs5 <6><7> rsfs7          %
  <8><9><10><10.95><12><14.4><17.28><20.74><24.88> rsfs10}{}     %
\DeclareMathAlphabet{\mathfs}{U}{rsfs}{m}{n}                     %
\newcommand{\mfs}[1]{\mathfs {#1}}                               %

\newcommand{\ba}{\nopagebreak[3]\begin{eqnarray}}
\newcommand{\ea}{\end{eqnarray}}
\newcommand{\bii}{\begin{itemize}}
\newcommand{\eii}{\end{itemize}}
\newcommand{\nn}{\nonumber}

\newcommand{\sO}{{\mfs O}}

\newcommand{\f}{\frac}

\def \d{\delta}
\def \b{\beta}
\def \l{\ell}
\def \g{\gamma}
\def \e{\epsilon}
\def \lp{\l_p}
\def \j{\sqrt{j(j+1)}}

\def \lm{\lambda}

\def \k{\kappa}
\def \H{\mathcal{H}}
\def \P{\mathcal{P}}

\def \sj{s_j^{\star}}

\def \O{\Omega}

\def \h{\hat}

\def \({\left(}
\def \){\right)}
\def \[{\left[}
\def \]{\right]}

\begin{document}
\title{Energy spectrum of black holes: a new view}
\author{Abhishek Majhi}%
\email{abhim@imsc.res.in, abhishek.majhi@gmail.com}
\affiliation{The Institute of Mathematical Sciences\\4th Cross Street, Taramani Campus\\Chennai, India\\}%

\begin{abstract}
Energy of a black hole is usually quantized by invoking some area quantization scheme after expressing the energy in terms of the horizon area. However, in this approach one has to quantize the local and asymptotic energy of the black hole separately and the two results do not manifest any physical correspondence with each other. Here, as opposed to this practice, we find the unique energy  spectrum of black holes by adopting a top-down approach. The physical links among the underlying quantum theory, statistical mechanics and thermodynamics of the black hole horizon play the central role in determining the energy spectrum.  The energy spectrum that we obtain, explicitly reveals the correspondence between asymptotic and local observations through the presence of the surface gravity of the horizon as a parameter in the spectrum, rather than being expressed as a function of area and consequently getting quantized in the usual approach.  Thus, our result presents a new view as far as black hole energy quantization is concerned. The calculations are performed using the quantum geometric description of black hole horizons as laid down by loop quantum gravity. 
\end{abstract}

\maketitle
\section{Introduction}
Efforts to quantize black hole energy date back to the works of Bekenstein\cite{bekmass} and Mukhanov\cite{mukmass}. The idea was to express black hole energy as a function of its horizon area and then to apply a specific area quantization ansatz, motivated by quantum mechanical intuitions, leading to quantized energy. However, these works had no connection with any underlying theory of quantum gravity as there were none present at that time.
Now, based on the quantum theory of black hole horizon geometry\cite{qg1,qg2}, modeled as isolated horizon (IH)\cite{ih1,ih2,ih3,ih4,ih5},  two recent efforts have been put into work to quantize black hole energy viz. \cite{gp} and \cite{cr}. However, both these approaches have their respective pros and cons. To mention, the proposed model Hamiltonian\footnote{By `Hamiltonian' we shall mean `energy' in the present context.} operator for the quantum IH discussed in \cite{gp}, although, has been shown to be consistent with the associated first law of IH mechanics, but it is only defined within a local observer framework and the asymptotic observation  remains unclear. On the other hand,  in \cite{cr}, the asymptotic energy of a black hole is quantized without any attempt to relate the result to the local observer scenario. Also, neither it has been shown, nor it is manifest, whether the energy quantization consistently gives rise to the first law of IH mechanics.
The point of interest in the present context is that the methodology involved in all these efforts viz. \cite{bekmass,mukmass,gp,cr} has one thing in common which follows from our natural intuition i.e. to express the horizon energy as a function of its area and then to apply area quantization to yield the corresponding energy spectrum. The only difference is the use of different area quantization owing to availability or unavailability of the underlying quantum theory in the respective cases. 

In this paper, we adopt a counter intuitive top-down approach to the concerned problem which yields that the Hamiltonian operator associated with the quantum IH is given by 
\ba
\hat H_S~\equiv~ \f{\k_{IH}}{8\pi}~\hat A_S
\ea
where $\hat H_S$ and $\hat A_S$ denote the Hamiltonian operator and the area operator for the quantum IH respectively and $\k_{IH}$ is the associated surface gravity with the IH. The point of departure from the known literature, and what is new in this result, is the presence of the surface gravity of the horizon as a parametric scale factor rather than a function of area getting quantized thereafter. Since, surface gravity is defined only with respect to the choice of the class of observers characterized by the corresponding time evolution vector field\cite{ih3}, it explicitly manifests the observer dependence of the spectrum. This implies that the local and asymptotic observations of the energy spectrum of a black hole horizon is just related by a red-shift factor, which is as expected from our basic knowledge of general relativity\cite{wald}.

Let us debrief the structure of the paper in short. In Section (\ref{ihenergy}) we review the notion of energy associated with IH in order to explain its observer dependence and how it is manifested. In Section (\ref{assum}), we  assume a certain structure of the Hamiltonian operator for the quantum IH. The operator is structured so as to capture certain properties which are expected to be manifested by the Hamiltonian operator of the quantum IH, if it were known a priori. Section (\ref{search}) contains the detailed analysis which leads to the final result. Subsection(\ref{mpm}) contains the statistical mechanical analysis involving the entropy calculation of the quantum IH using the method of most probable distribution\cite{stat,stat1}. The area spectrum of the quantum IH being known in LQG, the analysis is performed in area ensemble in subsection (\ref{aensemble}). We calculate the most probable macrostate, the entropy and its infinitesimal variation. Having assumed a Hamiltonian (energy) operator and its spectrum for the quantum IH, the same analysis is performed in the usual energy ensemble in subsection(\ref{eensemble}). However, the results now contain unknown functions and parameters of the assumed Hamiltonian operator, which need to be fixed so as to find the unique form of the Hamiltonian operator for the quantum IH. The notion of temperature originating from the microscopic theory is introduced in this subsection. In subsection (\ref{dmatch}), we fix the functional form of the Hamiltonian operator by demanding that the most probable macrostate of the quantum IH found from the area and energy ensembles should match in detail. But an ambiguity in the scale factor continues to remain. In subsection (\ref{lawmatch}), the ambiguity in the scale factor is removed by demanding that the temperature obtained from the microscopic theory must be related to the surface gravity which appears in the classical first law of IH mechanics by the usual relation {\it a la} Hawking\cite{hawkrad}. Having determined the form of the Hamiltonian uniquely (i.e. absence of any ambiguities), in subsection (\ref{finalform}), we cast the Hamiltonian finally in a meaningful form. In subsection (\ref{sexplain}), the non-triviality in our result and how it differs from the results obtained earlier by usual energy quantization procedure has been explained. Finally, we end with a conclusion in section (\ref{concl}).

{\it Units and Dimensions :} Before going into the matters, let us spell out the units we will be using and dimensions of certain relevant quantities which will come up in course of our analysis. We shall consider $G=1,c=1$ and dimensionally we will have $[Mass] = [Time] = [Length]$. Hence, energy will be expressed in the units of Planck length$(\lp=\sqrt{G\hbar/c^3})$. Further, we shall consider the Boltzmann constant $k_B=1$, which in turn implies dimensionally $[Temperature] = [Length]$ and hence temperature will also be expressed in the units of $\lp$ instead of Kelvin. Also, we shall shortly encounter a parameter $\b$ which will be identified as the inverse of the {\it dimensionless} temperature i.e. $\b=\lp/T$. We shall drop the word ``dimensionless'' and address $\b^{-1}$ as ``temperature'' only.

\section{Isolated horizon : associated notion of energy}\label{ihenergy}
 IH is a three dimensional null inner boundary of four dimensional spacetime, which is topologically $S^2\times R$. It satisfies certain boundary conditions so as to capture all the physical properties associated with the event horizon of a black hole spacetime. However, unlike the notion of event horizon, the description of an IH neither  requires the knowledge of asymptotic structure of the spacetime, nor does it require stationarity in the exterior bulk. Matter and radiation can reside arbitrarily close to the IH. 
 
  Now, the relevant aspect of an IH in the present context is the associated notion of energy. The Hamiltonian evolution of the classical phase space of spacetime with IH as an inner boundary requires a first law of mechanics to be satisfied on the IH, which can be written down as
 \ba
 \d E^t_{IH}=\f{\k^t_{IH}}{8\pi}\d A+\text{work terms}\label{flaw}
 \ea
where $t$ denotes a family of equivalent class of time evolution vector fields signifying a preferred foliation of the spacetime for the required Hamiltonian description. $E^t_{IH}$ and $\k^t_{IH}$ are the energy and surface gravity associated with the IH corresponding to a particular class of time evolution vector field $t$. $A$ is the classical area of the IH. Choosing this $t$ accordingly gives rise to the notion of asymptotic (ADM) energy of the IH\cite{ih3} or the energy observed by the class of local observers\cite{fgp}. In other words, a preferred foliation of the spacetime corresponding to a specific $t$ physically signifies the viewpoint of a  specific class of observers. Hence, the notion of energy associated with the black hole horizon is observer dependent.
 
   Although the first law of IH mechanics uniquely provides a notion of black hole energy for both local and asymptotic observers, we do not have an energy spectrum for black holes which explains both the local and asymptotic viewpoints in a single framework. Regarding this, one should particularly note that the surface gravity carries the burden of observer dependence in the dynamical law. So it may be expected that in the quantum theory the energy spectrum of the black hole possibly manifest this observer dependence through the presence of this surface gravity as a parameter. As we proceed, we shall see that this is  indeed the case if the energy spectrum of the black hole has to consistently give rise to the first law of IH mechanics. It may be noted that we shall drop the superscript $t$ on $E_{IH}$ and $\k_{IH}$ henceforth and will not be considering the work terms as we shall work with black holes without any charge and angular momentum.

\section{Expected structure of the Hamiltonian operator}\label{assum}
In this section, we shall write down the general structure of an operator which captures the expected properties of the Hamiltonian operator for the quantum IH, if it were known a priori.   

The quantum geometry of a cross-section of an IH is realized as a sphere with punctures made by the edges of the bulk spin-network graphs that describe the quantum geometry of space in LQG framework and each puncture is endowed with an SU(2) spin representation carried by the corresponding edge of the spin-network graph. Given a set of $N$ punctures on the quantum IH with spins $(j_1,\cdots, j_N)$, the associated Hilbert space is given by
\ba
\H_{S}(j_1,\cdots, j_N)=\text{Inv} \left(\otimes_{l=1}^N \H_{j_l}\right)
\ea
where $\f{1}{2}\leq j_l\leq\f{k}{2}\forall l\in[1,N] $, $k\equiv A_{IH}/4\pi\g\lp^2\in I$, $S$ represents the horizon, $A_{IH}$ is the classical area of the IH, $\g$ is the Barbero-Immirzi parameter and $\lp$ is the Planck length\cite{qg1,qg2,km98}. To be consistent with the above structure of the Hilbert space, the structure of the model Hamiltonian operator and its action on a state represented by a set of spins $(j_1,\cdots,j_N)$ may be {\it assumed} to be given by 
\ba
\h H_S|j_1,\cdots,j_N\rangle&=& \left(\h H_{1}\otimes \h I_{2}\otimes\cdots\otimes I_{N}+\h I_{1}\otimes \h H_{2}\otimes\right.\nn\\ 
&&~~~~~~\left.
\cdots\otimes \h I_{N}+\cdots+\h I_{1}\otimes \h I_{2}\otimes\cdots\otimes \h H_{N}\right)|j_1,\cdots,j_N\rangle\nn\\
&&~~~~\nn\\
&=&\lp \left[f(\h A_{1})\otimes \h I_{2}\otimes\cdots\otimes I_{N}+\h I_{1}\otimes f(\h A_{2})\otimes\right.\nn\\ 
&&~~~~~~\left.
\cdots\otimes \h I_{N}+\cdots+\h I_{1}\otimes \h I_{2}\otimes\cdots\otimes f(\h A_{N})\right]|j_1,\cdots,j_N\rangle\nn\\
&&~~~~\nn\\
&=&\lp\sum_{l=1}^{N}\e_{j_l}|j_1,\cdots,j_N\rangle
\label{ham}
\ea
where $f(\hat A)$ is a function of the dimensionless area operator acting on a specific single puncture with spin $j$ i.e. $\hat A|j\rangle= 8\pi\g\j|j\rangle$ \cite{qg1,qg2} (scaled by $\lp^2$ to render it dimensionless) and $\e_{j_l}=f(a_{j_l})$ where $a_j=8\pi\g\j$. The proposed operator structure guarantees certain properties which the Hamiltonian operator is expected to hold viz. the commutativity with the area operator, gauge invariance and self-adjointness (see \cite{am3} for explanation). Now, the crucial task is to find a suitable form of $f$ which will define the Hamiltonian uniquely. We did not incorporate any interaction terms in the assumed structure of Hamiltonian operator because of the fact that the punctures of the quantum IH are non-interacting by the very prescription of the theory itself\cite{qg1,qg2}.

One may note in course of the forthcoming analysis that the assumption of single puncture contribution to the Hamiltonian to be of the form $f(\hat A)$ will not play any role at all in arriving at the final result. In fact $\e_j$ will be determined to be proportional to $a_j$ implying $f(\hat A)\propto \hat A$ as the one and only possibility. However, to have a logical explanation and motivation behind the assumption, as already explained with proper reference, we needed to start with a specific prescription. Otherwise, it would have sufficed to begin with only the last line of eq.(\ref{ham}) and omitting the first two. Unfortunately, that would only look like a start with some blind guess, which it is not. 

\section{Search for the unique Hamiltonian operator}\label{search}

In this section we present the technical analysis to find a unique Hamiltonian operator for the quantum IH and the plan of progress can be briefly sketched as follows. We begin with finding the entropically favourable most probable macrostate which represents the quantum state of the quantum IH most closely representing its classical nature, which is in accord with the basic postulates and the necessity for the applicability of equilibrium statistical mechanics\cite{stat,stat1}. Demanding certain consistency conditions to hold, the model will be fixed uniquely.

\subsection{The most probable macrostate of the quantum IH}\label{mpm}
{\it The Hilbert space :} The full Hilbert space of an IH can be written as follows  
\ba
\H_{S}^{k}=\bigoplus_{\{\P\}}\text{Inv} \left(\otimes_{l=1}^N \H_{j_l}\right)
\ea
where ${\{\P\}}\equiv {N ; \f{1}{2}\leq j_l\leq\f{k}{2}\forall l\in[1,N]} $, $k\equiv A_{IH}/4\pi\g\lp^2\in I$, $S$ represents the horizon, $A_{IH}$ is the classical area of the IH, $\g$ is the Barbero-Immirzi parameter and $\lp$ is the Planck length.

{\it The microstate count :} In the large $k$ limit (which holds for $A_{IH}\gg\sO(\lp^2)$), the number of physical microstates corresponding to a macrostate represented by a spin configuration $\{s_j\}$ is approximately given by\footnote{For an explicit calculation showing how this formula can be deduced from the SU(2) Chern-Simons state counting see \cite{sigma,ampm2}.}  
\ba
\Omega[\{s_j\}]\simeq N!\prod_j[(2j+1)^{s_j}/s_j!]\label{count}
\ea
where $N:=\sum_js_j$ represents the total number of punctures associated for any configuration $\{s_j\}$\footnote{It should be noted that unlike many instances (e.g. \cite{gp,am3}) in current literature on quantum IH, $N$ will not be fixed a priori to define the ensemble. Readers are referred to the {\it Appendix} to find the justification of this statement. }. Here $\{s_j\}$ represents a set of punctures among which $s_j$ number of punctures carry spin $j$, while $j$ runs from $1/2$ to $k/2$.

{\it Entropy :} Considering Boltzmann formula for entropy and setting the Boltzmann constant to unity, the entropy of the IH is given by
\ba
S=\log \[\sum_{\{s_j\}} \O[\{s_j\}]\]
\ea
where the argument of the logarithm represents the total number of microstates arising from all possible spin configurations, the sum being over all possible spin configurations  constrained by the definition of the ensemble. {\it Assuming} that the basic postulates of equilibrium statistical mechanics are valid in present scenario of quantum IH, there is one most probable configuration whose corresponding number of microstates is overwhelmingly large compared to any other configuration (see e.g.\cite{stat,stat1}) such that the entropy of the IH can be approximately given by the entropy of the most probable configuration alone i.e.  
\ba
S&=&\log \[\sum_{\{s_j\}} \O[\{s_j\}]\]\nn\\
&=&\log \O[\{\sj\}] + \text{contributions from the sub-dominant configurations}\nn\\
&\simeq&\log \O[\{\sj\}]\label{defent}
\ea
Hence, the most probable distribution is the one which extremizes the entropy corresponding to a spin configuration, subject to the constraint which defines the ensemble. However, we have now two options to proceed. Firstly, the area of an IH is constant and since we have the knowledge of the area spectrum of quantum IH, we can work in the fixed area ensemble. Secondly, since we have now a Hamiltonian operator and hence its spectrum, we can also work in the fixed energy ensemble because IH does not allow matter and radiation to cross and thus has a fixed energy. 

\subsubsection{The area ensemble}\label{aensemble}
The action of the area operator for the quantum IH on a state represented by a spin configuration $\{s_j\}$ is given by 
\ba
\hat A_S|\{s_j\}\rangle&=&8\pi\g\lp^2\sum_{j}s_j\j ~|\{s_j\}\rangle\label{aop}
\ea
Now, we shall find the most probable spin distribution for a quantum IH by working in the area ensemble. So, we define an ensemble of quantum IHs by fixing their quantum areas to lie within the window $A_{IH}\pm\sO(\lp^2)$. Considering the fact that we are working with large IH having $A_{IH}\gg\sO(\lp^2)$, the spin configurations that are of interest in this context satisfy the  constraint
\ba
C_{area} : \sum_js_j\j- A_{IH}/8\pi\g\lp^2=0 \label{carea}
\ea
Extremization of the entropy $\log[\{s_j\}]$  corresponding to a spin configuration $\{s_j\}$ subject to constraint $C_{area}$, mathematically written as
\ba
\bar{\d} \log\Omega[\{s_j\}]-\lm~\bar{\d} C_{area}=0~~~,\nn
\ea
yields the distribution for the dominant spin configuration $\{\sj\}$ to be 
\ba
\sj=N_0(2j+1)\exp-\lm \j\label{mpd1}					
\ea
where $N_0:=\sum_j\sj$ is the total number of punctures corresponding to the most probable spin configuration and $\bar{\d}$ represents arbitrary variation with respect to the variable $s_j$. 

$\lm$ is a Lagrange multiplier that gets determined by the equation
\ba
\sum_j(2j+1)\exp-\lm \j=1\label{es}
\ea 
which is obtained by summing over $j$ on both sides of eq.(\ref{mpd1}) and using back the definition of $N_0$. Let us suppose its value comes out to be $\lm=\lm_0$(say)\footnote{The exact value of $\lm_0$ is not relevant in the present context. However, an estimate of the number can be found in \cite{am2,ampm2}.}. The entropy of the IH, as given by eq.(\ref{defent}), comes out to be 
\ba
S={\lm_0 A_{IH}}/{8\pi\g\lp^2}\label{enta}
\ea
where one has to use eq.(\ref{count}) followed by the application of Stirling's approximation considering the fact that $\sj\gg 1$ and finally using eq.(\ref{carea}) satisfied by $\sj$. 

Taking the differential of both sides of eq.(\ref{enta}) we obtain
\ba
\d S=\f{\lm_0}{8\pi\g}~\d( A_{IH}/\lp^2)\label{difarea}
\ea
which is the equation that we shall use further.

\subsubsection{The energy ensemble}\label{eensemble}
The action of the Hamiltonian operator for the quantum IH, as proposed in eq.(\ref{ham}), on a state represented by a spin configuration $\{s_j\}$ is given by 
\ba
\hat H_S|\{s_j\}\rangle&=&\lp\sum_{j} s_j\e_j ~|\{s_j\}\rangle\label{eop}
\ea
Now, we consider an ensemble of quantum IHs defined by fixing the energy in the window $E_{IH}\pm\sO(\lp)$. Considering the fact that we are working with massive IH having $E_{IH}\gg\sO(\lp)$, the spin configurations that are of interest in this context satisfy the  constraint
\ba
C_{energy} : \sum_js_j\e_j- E_{IH}/\lp=0 \label{cenergy}
\ea
Extremization of the entropy for a spin configuration subject to constraint $C_{energy}$, mathematically written as
\ba
\bar{\d} \log\Omega[\{s_j\}]-\b~\bar{\d} C_{energy}=0~~~,\nn
\ea
yields the distribution for the dominant spin configuration $\{\sj\}$ to be 
\ba
\sj=N_0(2j+1)\exp-\b\e_j\label{mpd2}					
\ea
where $N_0:=\sum_j\sj$ is the total number of punctures corresponding to the most probable spin configuration and $\bar{\d}$ represents arbitrary variation with respect to the variable $s_j$.

$\b$ is a Lagrange multiplier which satisfies the equation
\ba
\sum_j(2j+1)\exp-\b \e_j=1\label{ese}
\ea 
which is obtained by summing over $j$ on both sides of eq.(\ref{mpd2}) and using back the definition of $N_0$. The entropy of the IH, as given by eq.(\ref{defent}), comes out to be
\ba
S=\b E_{IH}/\lp\label{ente}
\ea
where one has to use eq.(\ref{count}) followed by the application of Stirling's approximation considering the fact that $\sj\gg 1$ and finally using eq.(\ref{cenergy}) satisfied by $\sj$.

Taking the differential of both the sides of eq.(\ref{ente}) yields
\ba
\d S=\b ~\d (E_{IH}/\lp)\label{difenergy}
\ea
As usual like other thermodynamic systems , $\b=\partial S/\partial (E_{IH}/\lp)$ can be interpreted as the temperature associated with the quantum IH.

\subsection{Detailed matching of the area and the energy ensembles}\label{dmatch}
There should be no dispute about the fact that we have studied the same physical system in two differently defined ensembles. Hence, the distribution function for the most probable configuration, which gives the physical probability distribution of the spins of the quantum IH in equilibrium, must be independent of whether we deal with area or energy ensemble. This implies that  we must have 
\ba
\e_j=\f{\lm_0}{\b} \j	\label{ej}				
\ea
which we obtain by comparing the most probable distributions in the two ensembles given by eq.(\ref{mpd1}) and eq.(\ref{mpd2}) and then putting $\lm=\lm_0$. Now, we shall carry out some  further consistency checks with the results obtained in the area and energy ensembles of quantum IHs.

\subsection{Consistency with the first law of IH mechanics}\label{lawmatch}
Eq.(\ref{difarea}) and eq.(\ref{difenergy}) both represent the change in entropy due to an infinitesimal shift in the equilibrium of the quantum IH. Hence, equating the two expressions we obtain
\ba
\d E_{IH}= \f{\lm_0}{8\pi\g\b\lp}~\d A_{IH}
\ea
which relates the infinitesimal change in energy to the infinitesimal change in area due to the shift in equilibrium of the quantum IH calculated by statistical mechanical methods. On the other hand we have the first law of IH mechanics\cite{ih3}
\ba
\d E_{IH}= \f{\k_{IH}}{8\pi}~\d A_{IH}
\ea
representing the same physical process. Demanding the compatibility of the two results and hence comparing the above two results, we obtain
\ba
\f{1}{\b}=\f{\g\lp}{\lm_0}\k_{IH}\label{temsg}
\ea
Now, only for the choice $\g=\lm_0/2\pi$ we have $\b^{-1}=\lp\k_{IH}/2\pi $ and then $\b$ can be interpreted as the inverse temperature of the quantum IH obtained from statistical mechanical methods having the usual relation to surface gravity. Also, then we can cast the first law of IH mechanics as the first law of thermodynamics for quantum IH viz.~
$\d S=\b~\d (E_{IH}/\lp)$. This concludes all the consistency checks between area and energy ensemble calculations aimed towards finding the unique Hamiltonian operator for the quantum IH. 

\subsection{Casting the Hamiltonian in a meaningful form}\label{finalform}
After going through the above consistency checks, now we have a suitable form of the Hamiltonian operator for the quantum IH which is designed to be consistent with the underlying statistical mechanical framework and the classical notion of energy associated with the IH defined for all class of observers. We shall cast it in a meaningful form as follows. At first, using eq.(\ref{ej}) back in eq.(\ref{eop}), we can write
\ba
\hat H_S|\{s_j\}\rangle &=& \lp\f{\lm_0}{\b} \sum_{l=1}^N s_j\j~|\{s_j\}\rangle\label{ham1}
\ea
which, by using eq.(\ref{temsg}), can be recast as 
\ba
\hat H_S|\{s_j\}\rangle &=& \lp^2\g\k_{IH}\sum_{j} s_j\j~|\{s_j\}\rangle\label{ham2}
\ea
Finally, using the action of the area operator on the quantum IH states given by eq.(\ref{aop}), the Hamiltonian can be written in a meaningful form as follows 
\ba
\hat H_S|\{s_j\}\rangle &=& \f{\k_{IH}}{8\pi}\hat A_S~|\{s_j\}\rangle\label{ham3}
\ea
In the process, we have restrored $\g$, which takes value $\lm_0/2\pi$. It may be noted that $\k_{IH}$ carries a dimension of length inverse in the units considered here. 


\subsection{Observer dependence and surface gravity}\label{sexplain}
For each and every choice of the time evolution vector field there is a first law (and hence energy) associated with the IH\cite{ih3}. Hence, the notion of energy of an IH is observer dependent and so should be the Hamiltonian operator and its spectrum. The choice of the time evolution vector field compatible with an asymptotic observer yields the ADM mass (see \cite{ih3}), and the choice compatible with the Rindler observer\cite{fgp}(see also \cite{bia1,bia2,smo,carlip1,carlip2,pady}) gives rise to the Rindler energy\cite{suss}. Area of the cross section of an IH is the only quantity which is intrinsically defined on the horizon and hence independent of any observer.  Surface gravity of the horizon manifests the crucial observer dependence in the first law of IH mechanics\cite{ih3}. This observer dependence must be manifested at the quantum level also and this is what has been shown here. The Hamiltonian operator for the quantum IH obtained here has this particular feature i.e. observer dependence. This is due to the algebraic structure of the operator which is of the form 
\ba
\hat H_S\equiv \cal R \times  \hat O
\ea
where $\cal R$ is the scale factor and $\hat {\cal O}$ is the operator part which actually acts on the states of the quantum IH and is intrinsically defined on the IH i.e. the area operator $\hat A_S$. This mathematical structure of the Hamiltonian can not be trivially guessed or justified if we would have  tried to quantize the energy of a known black hole expressed as a function of area. The obvious intuition would have been to lift the function $E=E(A)$ to the status of an operator $\hat H\equiv E(\hat A)$ (see for instance \cite{cr}). However, in that case it is not at all clear how one can relate the asymptotic result with the local observer scenario.

\section{Conclusion and outlook}\label{concl}
To conclude, from the above simple looking counter intuitive analysis, we gained the insight that the usual notion of energy quantization has to be modified so as to manifest the observer dependence alike the classical energy and then only it will be possible to explain a consistent correspondence between the observations of local and asymptotic observers. For example, one can express the ADM mass of a Schwarzschild black hole either as $M=\sqrt A/4\sqrt\pi$ or as $M=(\k/4\pi) A$. The first expression, being a function of area (which has an intrinsic geometric definition on the horizon), gives an impression that the energy is also an intrinsically defined quantity on the horizon, which is of course not true. Whereas the second expression manifests the observer dependence of the energy through the association of surface gravity of the horizon which is observer dependent. So, it is quite obvious that the second viewpoint to express the energy of a black hole is physically more justified. This indicates that the energy spectrum  of the quantum IH should have a similar structure manifesting the observer dependence and this is what we have obtained by adopting a counter intuitive approach in this work.  

Now, having such an energy operator at hand , it will be interesting to explore the thermodynamic stability analysis of black holes with this Hamiltonian operator. Calculating the surface gravity for a local observer and obtaining the corresponding energy spectrum, it has been shown  by a thermodynamic analysis that the horizon is thermodynamically stable\cite{gp}. However, the case for asymptotic observers remains to be explored using the energy spectrum discussed here. We shall report this issue elsewhere.

Finally, since this work has been performed within the realms of LQG, it is worth pointing out some future research directions related to this context, the hints to which are apparent from the present analysis. To perform the analysis, we have adopted a very awkward viewpoint in this work by proceeding towards the classical regime through quantum statistical mechanics. This is heavily based on the postulates of statistical mechanics rather than the quantum dynamics of the horizon. The reason behind this is the lack of our understanding of the coherent state description of IH. But, this is still a major problem of LQG itself. Although different approaches and viewpoints exist in literature (e.g. \cite{bahr1,bahr2,ori1,ori2,adas1, fri1}), nothing concrete has been done. However, in the realm of equilibrium statistical mechanics, given that the thermodynamics of a system is a theory of classical variables which emerges out of the statistical expected values of the corresponding observables of the  underlying quantum theory in the manner that has been investigated in the current piece of work, it is tempting to speculate that there might be a relation between the most probable macrostate and coherent states of the quantum IH. While one tries to approach the problem of coherent states of quantum IH, this speculation can serve as a good lead to what can be upcoming. Further, if this is proven to be true, the idea might be applicable beyond quantum IH framework to more generic quantum systems with similar Hilbert space structure.

\section{Appendix}

\subsection{Necessity of the commutativity with the area operator}

The reason behind seeking the commutation between the Hamiltonian and area operators of the quantum IH (i.e. $\left[\h H_S ,\h A_S\right]\equiv\h 0$) is that, it will in turn lead to the result that expectation value of the area operator of the quantum IH is a constant of motion i.e. in the correspondence limit the classical IH has {\it constant area}\cite{ih1,ih2,ih3,ih4,ih5}.  Mathematically, if there is some evolution parameter $\xi$ which parametrizes the evolution of the quantum IH, then it is evident that $\f{d}{d\xi}\langle\h A_S\rangle=\f{i}{\hbar}\langle\left[\h H_S ,\h A_S\right]\rangle=0$. 
  
However, there are two questionable issues regarding the requirement of commutativity of the Hamiltonian and the area operator for the quantum IH. It may be noted that the result  --  the classical area of the horizon remains constant, can be obtained in the correspondence limit in two other ways without the requirement of the commutativity of the Hamiltonian with the area operator. However, those two possibilities can be ruled out as follows:
\begin{itemize}
\item The first possibility is that $\hat H_S|j_1,\cdots,j_N\rangle=0$ for arbitrary sets of spins on the quantum IH and in that case, expectation value of the commutator will vanish even if the Hamiltonian does not commute with the area operator. This is tantamount to saying that the model Hamiltonian operator is nothing other than a scaled quantum version of CS Hamiltonian constraint $H_{CS}\approx 0$. However, this possibility can be discarded due to the following reason. It should be noted that, in the present scenario, the CS theory is on a three dimensional boundary of a four dimensional spacetime(bulk) and hence, it is not a free CS theory on a three-fold.  There is an interplay between the bulk and the boundary.  It has been shown at the classical level that there is a notion of non-zero energy associated with an IH\cite{ih3}. Thus, at the quantum level there has to be an energy spectrum associated with the quantum IH.

 
\item The second possibility is that the commutator does not vanish, but the expectation value vanishes  in the limit $\hbar\to 0$. But this will mean that the Hamiltonian operator contains some interaction terms rendering it with off-diagonal terms which prevents it to 
commute with the diagonal area operator. This opposes the fact that the punctures are non-interacting -- which follows from the quantum theory itself. Hence, this second possibility is also ruled out.
\end{itemize}

\subsection{No reason to fix $N$ {\it a priori}}
This paragraph provides an explanation to the fact pointed out in {\it footnote 3}. It is evident from the full Hilbert space structure of the quantum IH that the arbitrariness of $N$ only represents some sort of quantum geometric fluctuations underlying the classical structure of the IH. The variable $N$ neither represents any conserved quantity of the  theory (unlike mass, angular momentum and charge parameters which are calculable from the classical theory), nor there is any well defined number (of punctures) operator for the quantum IH. The quantum variable $N$ is an artifact of the LQG quantization procedure which is introduced in course of the point-splitting regularization of  the area operator for a two surface\cite{ashlew}. Hence, we do not find any plausible enough physical reason to consider the number of punctures to be fixed a priori in any related analysis (which was actually done in \cite{am3}). The idea of fixation of the number of punctures was basically imposed by hand\cite{gp}, which leads to some nontrivial and ad hoc consequences as far as entropy calculation is concerned \cite{ampm2,am2}.

\vspace{0.5cm}
{\bf Acknowledgements :}  The work is funded by the Department of Atomic Energy of the Government of India. I am grateful to Aloke Laddha, Ghanshyam Date, Romesh Kaul for offering interesting comments, and especially to Kalyan Rama for pointing out a crucial issue, during a seminar on the subject matter of this paper. I also thank Steven Carlip for pointing out the significance of this work in relation to the references\cite{carlip1,carlip2,pady}.

\end{document}